# Photonic molecules made of matched and mismatched microcavities: new functionalities of microlasers and optoelectronic components

Svetlana V. Boriskina[*1], Trevor M. Benson[2], Phillip Sewell[2]
[1]School of Radiophysics, V. Karazin Kharkov National University, Kharkov 61077, Ukraine
[2]G.Green Inst. for Electromagnetics Research, Univ. of Nottingham, Nottingham NG7 2RD, UK

## ABSTRACT

Photonic molecules, named by analogy with chemical molecules, are clusters of closely located electromagnetically interacting microcavities or "photonic atoms". As two or several microcavities are brought close together, their optical modes interact, and a rich spectrum of photonic molecule supermodes emerges, which depends both on geometrical and material properties of individual cavities and on their mutual interactions. Here, we discuss ways of controllable manipulation of photonic molecule supermodes, which improve or add new functionalities to microcavity-based optical components. We present several optimally-tuned photonic molecule designs for lowering thresholds of semiconductor microlasers, producing directional light emission, enhancing sensitivity of microcavity-based bio(chemical)sensors, and optimizing electromagnetic energy transfer around bends of coupled-cavity waveguides. Photonic molecules composed of identical microcavities as well as of microcavities with various degrees of size or material detuning are discussed. Microwave experiments on scaled photonic molecule structures are currently under way to confirm our theoretical predictions.

## 1. INTRODUCTION

When the concept of a photonic molecule was first introduced in 1998 [1], it provided a nice illustration of the parallels between behavior of photons in photonic atoms and electrons in either real or artificial atoms, while practical applications seemed elusive. In nine years that followed, the major international research effort in this field not only yielded insights in the new physical and optical phenomena observed in photonic molecules, but also brought rapid advances in their practical applications as semiconductor microlasers, wavelength-selective optoelectronic filters and switches, coupled-cavity waveguides, slow-wave structures, biosensing platforms, etc.

Manufacturing techniques have also matured and now make possible fabrication of photonic molecules of various materials, cavity shapes, and molecule structural configurations. Experimental techniques of the analysis of photonic molecule geometrical and spectral properties are now also well-developed. Simple and complex photonic molecules whose characteristics have been analyzed theoretically and experimentally include square semiconductor photonic dots connected by material bridges [1], photonic-crystal defect cavity structures [2], side-coupled thin-microdisk resonators [3], and various 2-D and 3-D structures composed of spherical microcavities [4-7]. Relative ease of manufacture, improved performance (in comparison to isolated cavities) that photonic molecules offer together with possibility to custom-tailor their properties by tuning the molecule geometry pave the way for their use in many existing and emerging optoelectronic and quantum-optical devices. Furthermore, photonic molecules can be used as a valuable testbed for many-body theories and cavity quantum electrodynamics experiments.

In this paper, we give an overview of applications for which using coupled-cavity photonic molecules instead of single microcavities yields improved or completely new functionalities. We compare properties of photonic molecules composed of identical microcavities with those of photonic molecules demonstrating either slight or significant size disorder. As it will be shown in the following sections, microcavities size mismatch should not always be considered as a disadvantage and may in fact be used for designing novel structures with enhanced performance. Unique opportunities offered by photonic molecules composed of microcavities of different types will also be discussed.

---

[*] SBoriskina@gmail.com; phone 380 57 7040373; fax 1 831 3087657; geocities.com/sveta2004

## 2. MODE SPLITTING IN DOUBLE-CAVITY PHOTONIC MOLECULES

Double-cavity structures have previously been used to investigate basic features of modes coupling in photonic molecules [1, 3]. Such structures are easy to analyze, while careful study of their properties provides a useful insight into the behavior of more complex coupled-cavity clusters. Here, we will briefly review the main features of double-cavity photonic molecules. Several scenarios of mode coupling in such molecules can occur, depending on the modal properties of individual cavities they are composed of. The first scenario is realized if the cavities support non-degenerate optical modes, such as e.g. defect modes in photonic crystal cavities shown in Fig 1a or fundamental modes in square pillar microcavities considered in [1]. If two such microcavities are electromagnetically coupled, non-degenerate individual cavity modes split into two non-degenerate bonding (symmetrical) and anti-bonding (anti-symmetrical) modes. For example, two neighboring supermodes are present in the optical spectrum of the photonic molecule consisting of two closely located defects in a photonic crystal lattice (Fig.1 b,c). The resonant wavelengths of these supermodes are shifted from that of the individual isolated cavity. The anti-bonding mode is red-shifted and the bonding mode is blue-shifted.

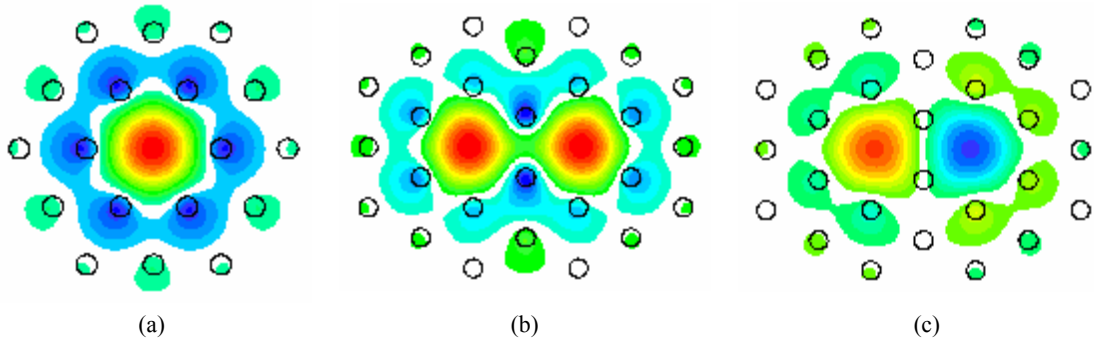

(a) (b) (c)

Fig. 1. The electric field profiles ($E_z$) of (a) a non-degenerate monopole mode in a hexagonal photonic crystal defect cavity ($\lambda$ = 9.054 µm, Q = 56.8); (b) bonding ($\lambda$ = 8.837 µm, Q = 90) and (c) anti-bonding ($\lambda$ = 9.324 µm, Q = 67.4) non-degenerate supermodes of a photonic molecule composed of two coupled hexagonal PC cavities. PC lattice parameters are as follows: $\varepsilon_{rods}$ = 8.41, $a$ = 0.6 µm, $d$ = 4.0 µm.

The photonic molecule optical spectrum has a more complex structure if individual microcavities support degenerate modes. For example, whispering-gallery modes supported by circular microdisk resonators may have very high quality factors yet are double-degenerate due to the symmetry of the microdisks [8-10]. Such modes are classified as $WG_{m,n}$ modes, $m$ being the azimuthal mode number, and $n$ the radial mode number. It has been shown, both theoretically and experimentally, that bringing microdisk resonators close together results in splitting of their degenerate WG-modes into two groups of nearly-degenerate photonic molecule supermodes (Fig. 2). In general, the number of nearly-degenerate modes in each group depends on the degeneracy of the individual cavity mode (e.g., two for a circular-disk photonic molecule [3]) and the number of groups is equal to the number of cavities. The mode with the highest Q-factor is the blue-shifted anti-bonding mode, which is anti-symmetrical with respect to both axes (Fig. 2a). WG modes in microspheres are multiple-degenerate, and thus double-microsphere molecules spectra reveal fine structure of supermodes with very close wavelengths [7]. Geometrical imperfections may push the supermode wavelengths further apart. Note that similar degenerate mode splitting occurs when the cavity is located close to a material interface or a bus waveguide [11].

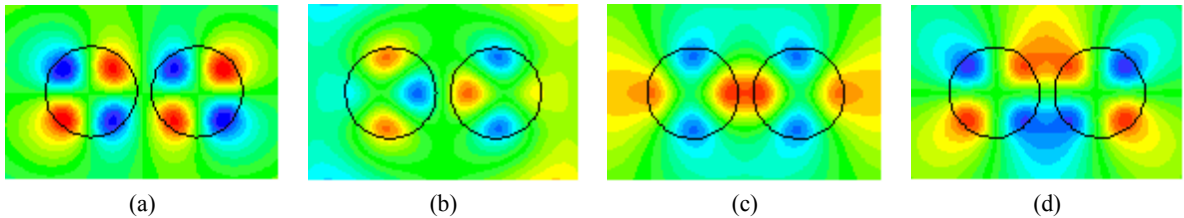

(a) (b) (c) (d)

Fig. 2. The electric field profiles ($E_z$) of four non-degenerate $WGH_{2,1}$-supermodes in a photonic molecule composed of two identical microdisks with $\varepsilon$ = 10.24+$i$10$^{-4}$, $a$ = 0.3 µm and $d$ = 0.7 µm. (a) OO: $\lambda$ = 1.567 µm, Q = 38.6; (b) EO: $\lambda$ = 1.6 µm, Q = 18.6; (c) EE: $\lambda$ = 1.661 µm, Q = 16.6; and (d) OE: $\lambda$ = 1.711 µm, Q = 29.6.

## 3. Q-FACTOR ENHANCEMENT OF PHOTONIC MOLECULE SUPERMODES

One of the most striking examples of the advantages offered by photonic molecules is a possibility of building quasi-single-mode structures with significantly (up to 50-fold) enhanced quality factors [12, 13]. This can be done by arranging identical WG-mode microcavities into high-symmetry photonic molecule structures such as triangles [12, 14], squares [12, 13], hexagons and circular loops [15] and by carefully tuning the inter-cavity coupling distances. Modal characteristics of a hexagonal-shaped photonic molecule are presented in Fig. 3a. An important feature of this figure is that Q-factors of some of the supermodes experience dramatic increase for certain values of the inter-cavity distances. This effect offers a way to design high-Q photonic molecule configurations. Here and thereafter, photonic molecule supermodes are classified by how they transform under the mirror operation with respect to the symmetry axes of the structure (e.g., OO supermode has odd symmetry with respect to all the symmetry axes of the hexagon).

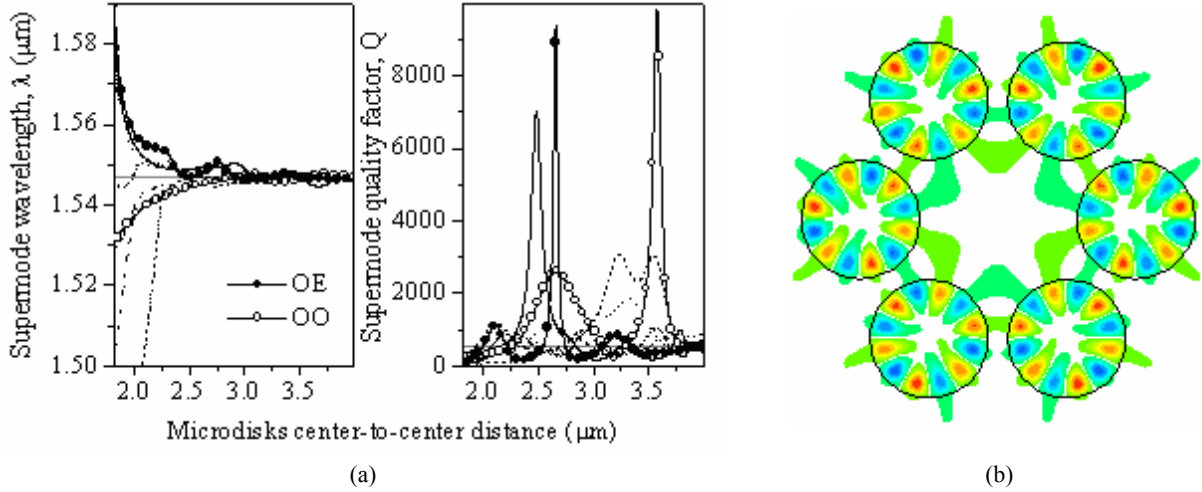

(a)            (b)

Fig. 3. (a) TE-polarized $WG_{6,1}$ mode wavelength splitting (left) and Q-factors change (right) in a hexagonal photonic molecule as a function of a distance from the microdisks centers to the molecule center. All the microdisks in the molecule have radii 0.9 μm and dielectric permittivity $\varepsilon=6.9169+i10^{-4}$. Corresponding single-disk characteristics are plotted for comparison (straight lines). (b) The magnetic field profile of the OE $WG_{6,1}$-supermode of the hexagonal photonic molecule.

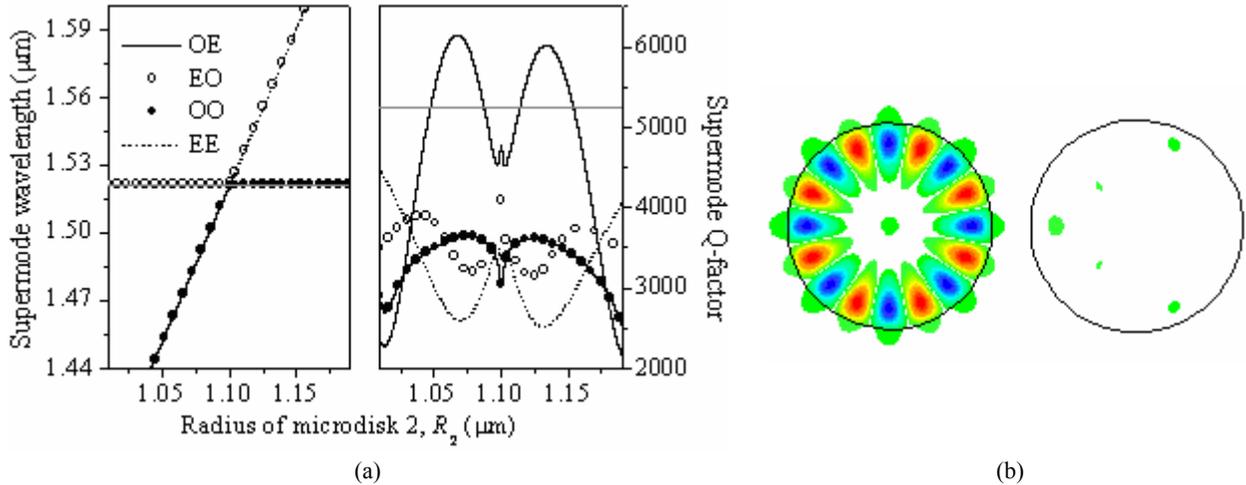

(a)            (b)

Fig. 4. (a) TE-polarized $WG_{8,1}$ mode wavelengths migration (left) and quality factors change (right) in a photonic molecule composed of two size-mismatched microdisks as a function of a radius of one of the disks. The other microdisk has a fixed radius of 1.1 μm. Both disks have dielectric permittivity $\varepsilon=6.9169+i10^{-4}$ and are separated by an airgap of 400 nm width. Corresponding single-disk characteristics are plotted for comparison (straight lines). (b) The magnetic field profile of the Q-enhanced OE-supermode of the size-mismatched double-disk photonic molecule ($R_2=1.065$ μm).

It should be noted that distortion of the molecule symmetry (e.g., cavities size disorder) may either reduce or completely suppress the effect of the Q-factor enhancement [14, 15]. It may be technologically challenging to fabricate completely identical microcavities. However, our studies reveal that, surprisingly, photonic molecules composed of size-mismatched microcavities can also be tuned to demonstrate enhanced performance [3, 16]. For example, Q-factor enhancement of one of supermodes can be achieved in a double-cavity photonic molecule composed of microdisks with slightly different radii (Fig. 4). Efficient mode coupling between microcavities with severe size mismatch is also possible [17, 18].

Note that in the photonic molecule configurations that yield enhancement of one of the supermodes all the other supermodes have much lower Q-factors (Fig. 3a, 4a). This means that optimally tuned photonic molecule structures have increased free spectral range in comparison to that of larger-radius individual microdisk resonators with comparable values of Q-factors [12]. Furthermore, as the high-Q modes of the photonic molecules are non-degenerate, such structures are expected to be less sensitive to fabrication imperfections that split WG-modes of single-cavity resonators causing appearance of parasitic peaks in their optical spectra.

## 4. ENHANCED SENSITIVITY TO ENVIRONMENTAL CHANGES

Photonic molecules can also be used as microcavity-based bio(chemical) sensors with enhanced sensitivity [13, 18]. Such sensors detect the resonant frequency shifts caused by the presence of a biochemical agent through the interaction of the evanescent field of the WG-mode outside microcavities with the analyte. As we have shown above, optimally-tuned photonic molecules have all the ingredients necessary to construct an efficient sensor. Supermodes with high Q-factors produce high-amplitude widely-spaced narrow features in the optical spectra of photonic molecules, making detection easier. Furthermore, photonic molecule supermodes, which are collective multi-cavity resonances, are more sensitive to changes of their nano-environment than the modes of single cavities owing to the presence of regions of high field intensity outside microcavities. One possible multi-cavity configuration that demonstrates simultaneously high Q-factor of a single non-degenerate OE-supermode and high sensitivity to the variations of the refractive index in the inter-cavity region [13] is shown in Fig. 5a,b. As illustrated in Fig. 5b, the sensitivity (determined by the slope of the lines) is enhanced for the sensor based on 4-cavity molecule and even further enhanced for the 16-cavity one.

On the other hand, combining record Q-factors achieved in WG-mode optical microcavities with exceptionally strong sub-wavelength field confinement observed in noble-metal nano-particles supporting surface plasmon resonances (Fig. 5c) is expected to result in the development of new classes of highly sensitive biosensors [19, 20]. To achieve the desired functionality of such hybrid devices, both micro- and nano-scale cavities have to be spectrally and geometrically engineered to maximize spatial and spectral overlap between their respective modal fields.

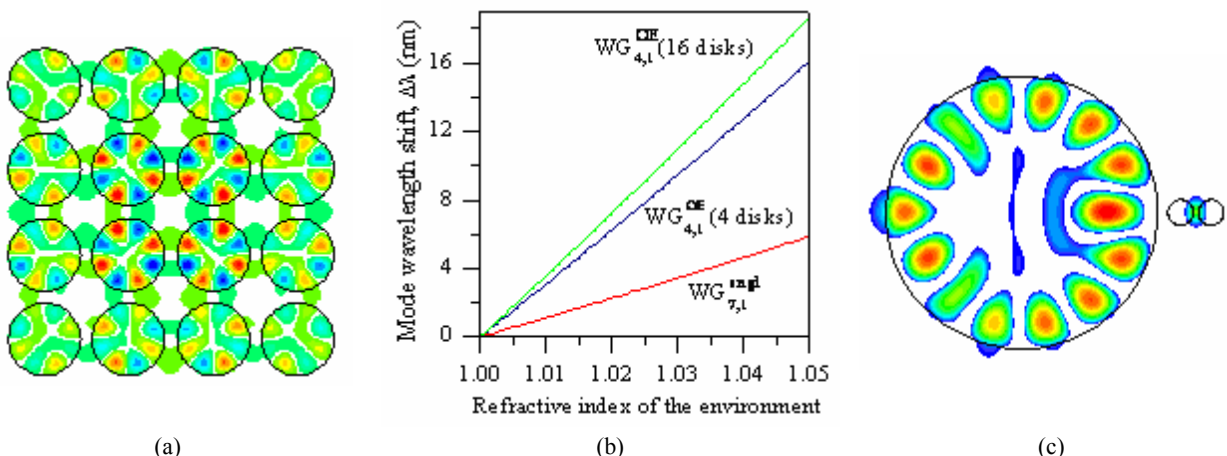

(a) (b) (c)

Fig. 5. (a) Near-field intensity portraits of the symmetry-enhanced OE mode of a square photonic molecule composed of 16 microdisks with radii of 0.67 μm; (b) Comparison of the sensitivity to changes in the refractive index of the surrounding of a single 2 μm diam microdisk operating on a $WG_{7,1}$ mode (λ = 1.531 μm, Q = 1660) with those of a 4-cavity molecule (λ = 1.566 μm, Q = 2768) and a 16-cavity molecule (λ = 1.578 μm, Q = 6656) operating on the symmetry-enhanced OE $WG_{4,1}$ modes; (c) Magnetic field distribution in a hybrid structure composed of evanescently coupled WG-mode microcavity and a chain of silver nanocavities supporting a localized surface plasmon mode.

## 5. DIRECTIONAL EMISSION FROM PHOTONIC MOLECULES

For many practical applications, it is highly desirable to obtain directional emission from microcavity structures. One of the disadvantages of conventional WG-mode circular or spherical microcavities is their multi-beam emission patterns. When several microcavities are electromagnetically coupled via their near-fields, this coupling also modifies their far-field emission characteristics. We have shown that by tuning the photonic molecule geometry, it is possible to design structures featuring only few narrow beams in the far-field pattern [20-22]. One of such structures is the photonic molecule composed of identical WG-mode microdisks arranged into an equilateral triangle shown in Fig. 6a. The EE-supermode supported by this structure yields directional emission pattern with three narrow beams (Fig. 6b).

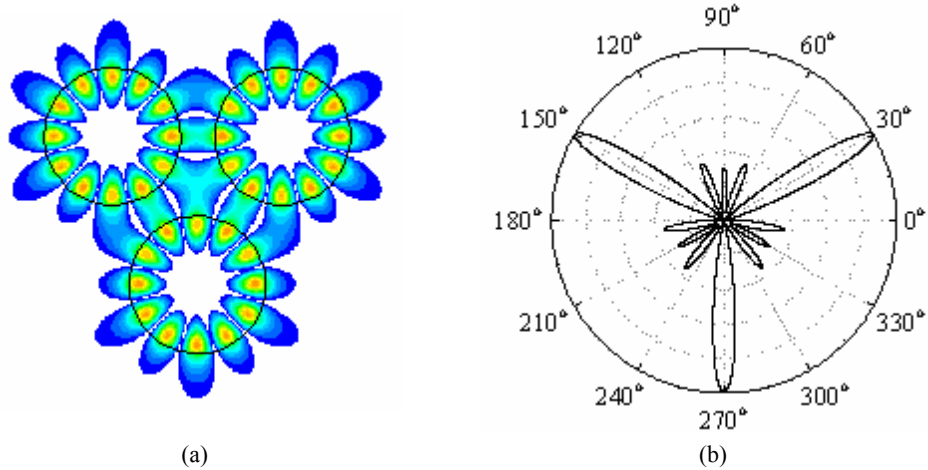

(a)            (b)

Fig. 6. Near-field intensity portrait (a) and directional far-field emission pattern (b) of the TM-polarized EE-WG$_{6,1}$ supermode of a triangular photonic molecule. Microdisks have 0.5 μm radii and dielectric permittivity ε = 10.24; the distance from the disks centers to the PM center is 0.72 μm.

Directional emission patterns can also be achieved in photonic molecules composed of size-mismatched microcavities, such as the structure shown in Fig. 7a. Here, tuning the distances between the individual cavities not only results in the increase of the photonic molecule supermode Q-factor but also yields directional emission pattern with well-defined directions of preferred light escape.

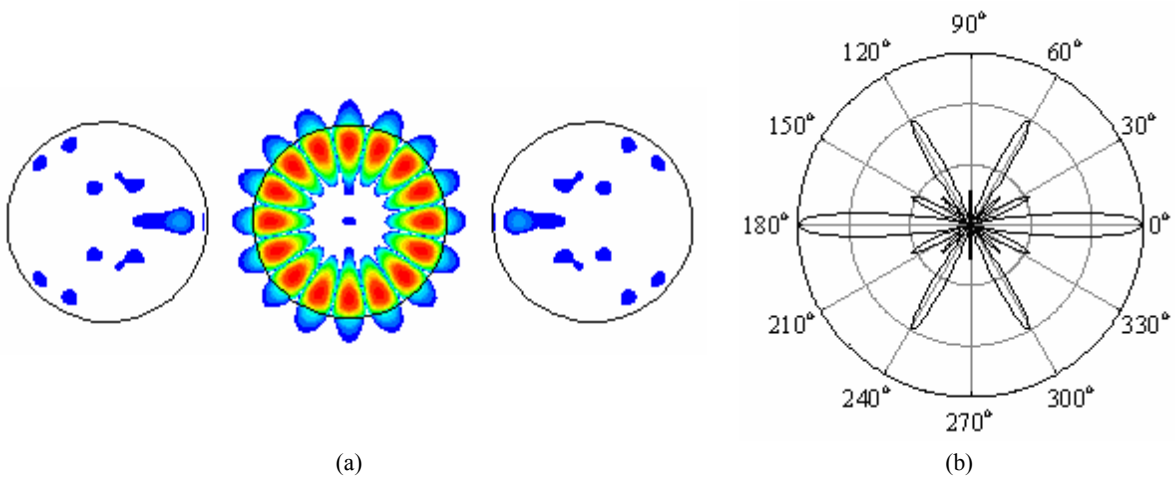

(a)            (b)

Fig. 7. Near-field intensity portrait (a) and directional far-field emission pattern (b) of the TE-polarized OE-WG$_{8,1}$ supermode of a size-mismatched photonic molecule. All three disks have dielectric permittivity ε=6.9169+$i$10$^{-4}$. The central disk of radius 1.065 μm is separated from the side disks of radii 1.1 μm by airgaps of 400 nm width.

## 6. REDUCTION OF CROW BEND LOSSES

Side-coupled microcavities also find use for the optical energy transfer and light slowing in the coupled-resonator optical waveguides [23, 24]. Optical modes spectra of infinite linear chains of microcavities have been extensively studied both theoretically and experimentally. Here, we focus on the potential application of CROWs for forming low-loss waveguide bends. This feature of CROWs has been predicted in the pioneering paper [23], where some basic rules for designing low-loss or even loss-less CROW bends have been envisioned. Our studies of finite-size curved CROW sections show that the recipe for forming low-loss CROW bends is more complicated than it was previously believed [25]. To illustrate our observations, Fig. 8a shows total power of the directional light beam scattered from the CROW section composed of seven side-coupled identical WG-mode microdisks in a narrow wavelength range and for several values of the CROW bend angle. The beam is grazing the rim of the left resonator in a chain. It is assumed that the right microdisk has material losses. Thus, minima in the plots on Fig. 8a correspond to the selected wavelengths at which the incident beam energy in transferred through CROW via coupled WG-modes (Fig. 8b) and is partially absorbed in the last resonator in the chain. The deeper is the minimum, the more efficient is the energy transfer. Clearly, the most efficient transfer can be achieved around CROW bends of 20 and 50 degrees.

Keeping in mind that a finite-size CROW section is in fact a photonic molecule of a special shape, we studied the change of the Q-factor of the CROW supermode that yields efficient energy transfer. As mentioned before, the supermodes form closely located doublets in the CROW spectrum. Thus, every minimum in Fig. 9a corresponds to two closely located supermodes. Plots in Fig. 9a reveal Q-factor enhancement of one of the modes in the doublet at the CROW bend angle of $22^o$. Similarly, the other mode is enhanced if the CROW bend angle is equal to $54^o$. Near-field portraits of these two modes are shown in Fig. 9b and indeed demonstrate low field intensity in the region of the bend and thus low bend radiation losses. To make possible efficient energy transfer along an arbitrarily bent CROW section, the size of the central disk can be tuned. As can be seen in Fig. 10, low bend losses for the angle of $90^o$ can be achieved in a CROW with the central disk having smaller radius than other disks. To confirm the theoretical predictions, microwave experiments on scaled coupled-cavity structures are being carried out at the University of Nottingham. Preliminary results for 4-cavity bent CROW section made of dielectric material C-Stock AK ($\varepsilon = 10$) are presented in Fig. 11.

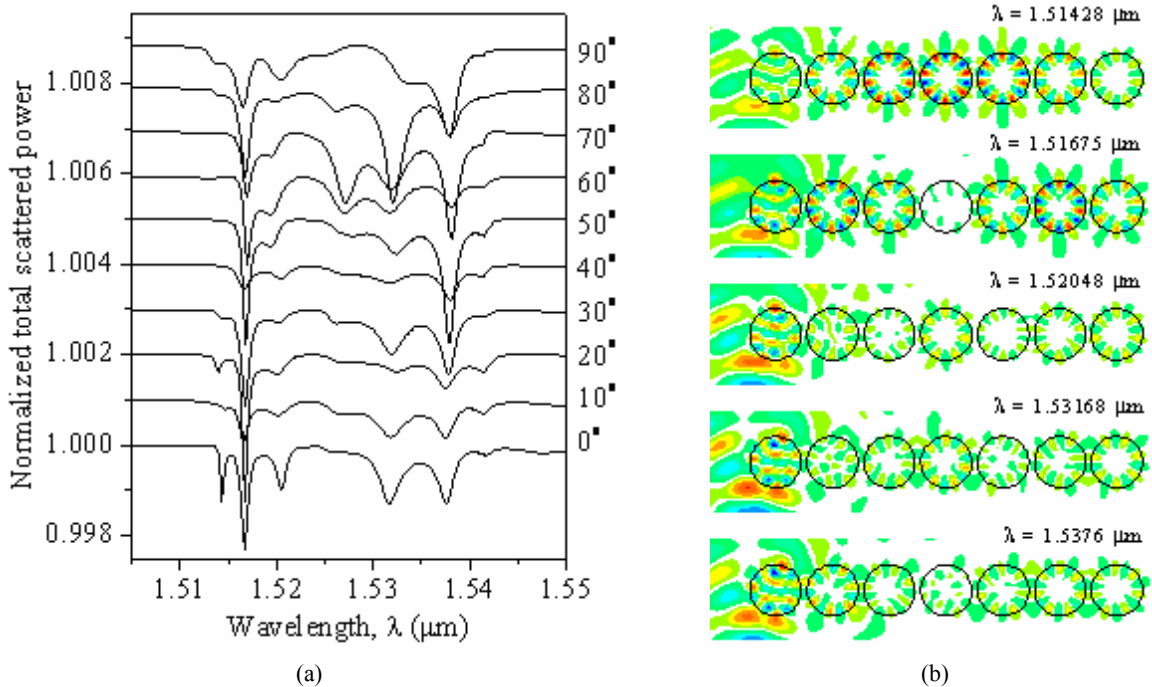

(a)        (b)

Fig. 8.(a) Normalized total scattered power of a TM-polarized CPS beam scattering from a CROW section composed of seven WG-mode microdisks as a function of the wavelength for several values of the CROW bend angle. All the microdisks in the CROW have radii of 0.9 μm, dielectric permittivity $\varepsilon=7$, and are separated by 200 nm airgaps. The beam is grazing left microcavity; it is assumed that right microcavity has absorption losses; (b) The electric field profiles in the straight CROW section at the wavelengths corresponding to the total scattered power minima observed in (a).

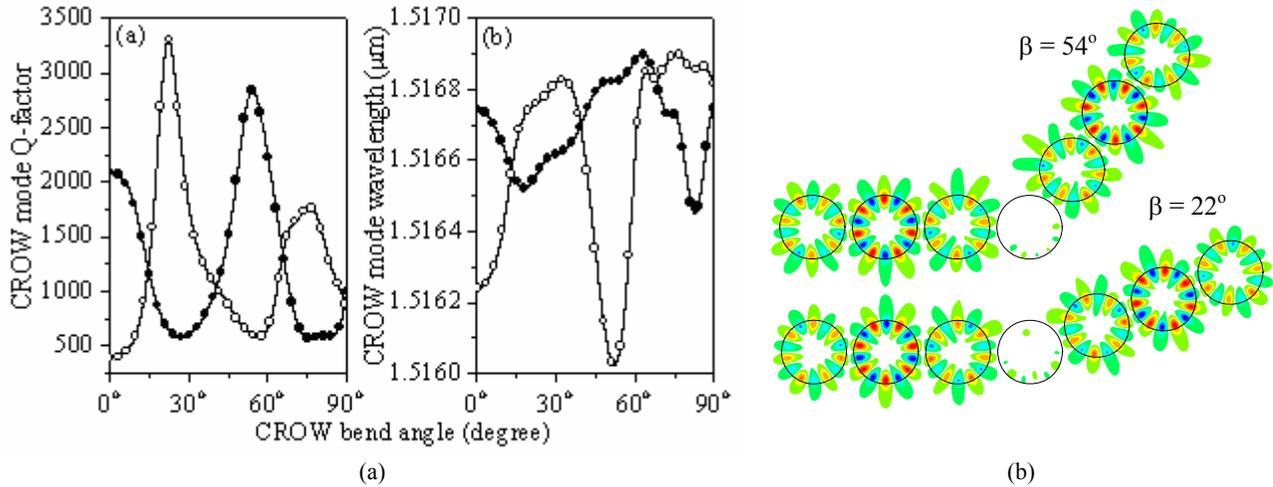

Fig. 9. (a) CROW quality factors change (left) and mode wavelengths migration (right) in a CROW with the same parameters as in Fig. 8 as a function of the CROW bend angle; (b) Near-field portraits of the CROW OO and EO anti-bonding supermodes with high values of Q-factors for the CROW bend angles of 54° and 22°.

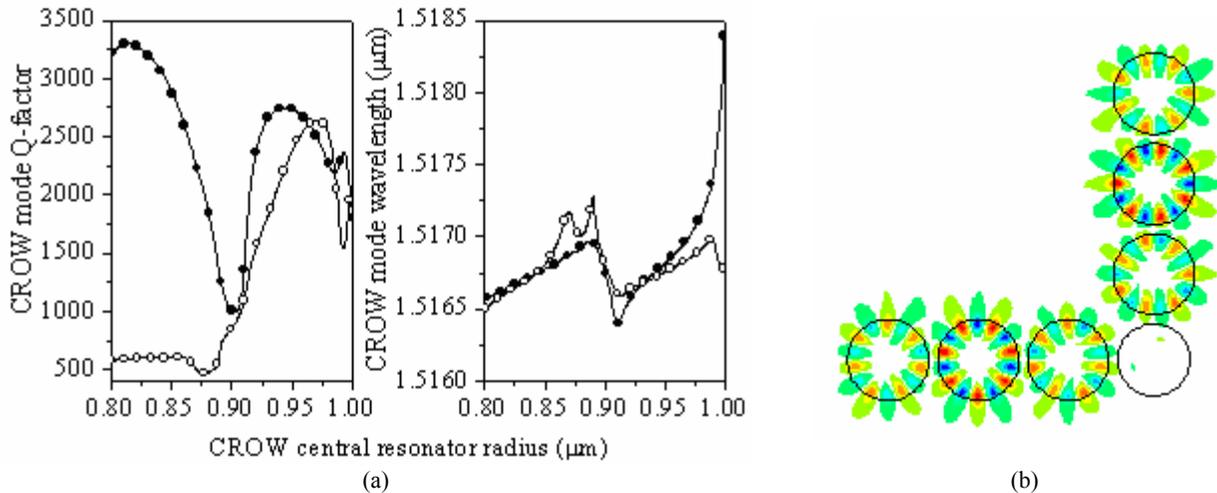

Fig. 10. CROW quality factors change (left) and mode wavelengths migration (right) in a CROW with the same parameters as in Fig. 8 and the bend angle of 90° as a function of the central resonator radius. (b) Near-field portrait of the EO anti-bonding supermode in the CROW with the central disk of 0.8 μm radius.

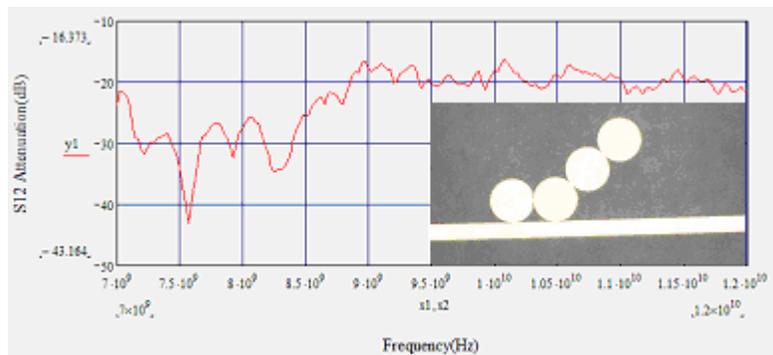

Fig. 11. Resonant spectra of the CROW made of 4 resonators of radii 1 cm and ε = 10 excited by a straight feed waveguide made of the same material. The CROW photograph is shown in the inset [27].

## 7. NANOJET-INDUCED MODES IN COUPLED-CAVITY CHAINS

Another light coupling mechanism that can be used to achieve energy transfer along a chain of side-coupled microcavities is nanojet-induced coupling [28-29]. Photonic nanojets are generated at the shadow-side of circular microcavities illuminated by a plane wave and propagate over several optical wavelengths without significant diffraction. It has previously been shown [28] that optical coupling and transport of nanojet-induced modes through a chain of microcavities is less efficient than the transport of WG-modes, which make use of the mechanism of the evanescent coupling. This is due to the radiative nature of photonic nanojets. Our results show that in some cases efficient transfer of a nanojet-induced mode can be achieved through coupled-cavity structures with significant size mismatch. For example, tuning the coupling distance between two identical miscrocavities shown in Fig. 12a does not result in efficient nanojet transfer through the second cavity. Changing their radii simultaneously did not yield any improvement of the transfer efficiency either. However, as can bee seen in Fig. 12b, efficient transfer can be achieved if the second cavity is approximately twice as large as the first one. This effect is observed for a wide range of values of the cavities radii ratio, and thus such structure is expected to have high fabrication tolerances.

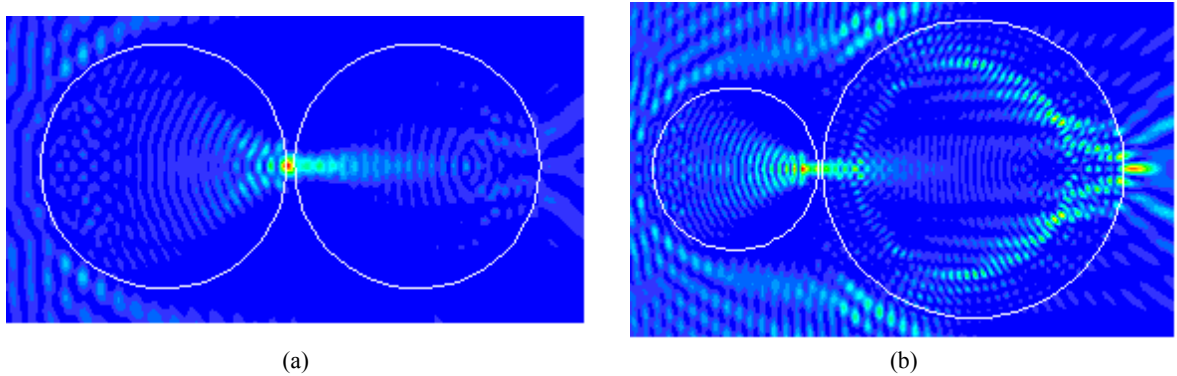

(a)                  (b)

Fig.12. Near-field intensity distributions in two side-coupled microcavities with a refractive index of 1.59 separated by an airgap of 200 nm: (a) both cavities have radii of 1.5 μm; (b) left cavity has radius 1.5 μm, right cavity has radius of 3.0 μm. TM-polarized plane wave of wavelength 425 nm propagates from left to right.

## 8. CONCLUSIONS

Novel functionalities offered by photonic molecules together with the ability to relatively easily fabricate a large variety of such 2D and 3D coupled-cavity structures paves the way for their use in various areas ranging from biotechnology to optoelectronics to quantum computing. Among the advantages offered by specially-designed photonic molecules are the following: (i) degenerate mode splitting and dramatic Q-factors enhancement of non-degenerate photonic molecule supermodes; (ii) increased free spectral ranges in comparison to those of isolated microcavities of larger radii and comparable Q-factors; (iii) directional emission patterns; (iv) enhanced sensitivity to the changes in the host medium; and (v) efficient energy transfer around waveguide bends. For the first time to our knowledge, we show that all these improved functionalities can be achieved not only in photonic molecules composed of identical microcavities but also in coupled-cavity structures with either slight or significant size mismatch. We have also explored ways to manipulate WG-mode cavity spectra and to control light out-coupling that can be achieved in hybrid photonic molecules via coupling of microcavity WG-modes to localized surface-plasmon modes of noble-metal nanoparticles.

## ACKNOWLEDGEMENTS

This work has been partially supported by the NATO Collaborative Linkage Grant CBP.NUKR.CLG 982430. We wish to thank Dr. Vasily Astratov for useful discussions and Ms Anika Arora for sharing her experimental results.